\documentclass{mn2e}
\usepackage{psfig}

\begin{document}

\date{\today} 

\title[X-ray flares]{X-ray flares from propagation instabilities in
  long Gamma-Ray Burst jets}

\author[Lazzati et al.]{D. Lazzati$^{1}$,
  C. H. Blackwell$^{1}$, B. J. Morsony$^{2}$ and M. C. Begelman$^{3,4}$\\
  $^{1}${Department of Physics, NC State University, 2401
    Stinson Drive, Raleigh, NC 27695-8202} \\
  $^{2}${Department of Astronomy, University of Wisconsin-Madison,
    5534 Sterling Hall, 475 N. Charter Street, Madison WI
    53706-1582}\\
  $^{3}${JILA, University
    of Colorado, 440 UCB, Boulder, CO 80309-0440}\\
  $^{4}${University of Colorado, Department of Astrophysical and
    Planetary Sciences, 391 UCB, Boulder, CO 80309-0391}}

\maketitle

\begin{abstract}
  We present a numerical simulation of a gamma-ray burst jet from a
  long-lasting engine in the core of a 16 solar mass Wolf-Rayet
  star. The engine is kept active for 6000~s with a luminosity that
  decays in time as a power-law with index $-5/3$. Even though there
  is no short time-scale variability in the injected engine
  luminosity, we find that the jet's kinetic luminosity outside the
  progenitor star is characterized by fluctuations with relatively
  short time scale. We analyze the temporal characteristics of those
  fluctuations and we find that they are consistent with the
  properties of observed flares in X-ray afterglows. The peak to
  continuum flux ratio of the flares in the simulation is consistent
  with some, but not all, the observed flares. We propose that
  propagation instabilities, rather than variability in the engine
  luminosity, are responsible for the X-ray flares with moderate
  contrast. Strong flares such as the one detected in GRB~050502B,
  instead, cannot be reproduced by this model and require strong
  variability in the engine activity.
\end{abstract}


\section{Introduction}

One of the most surprising discoveries of the Swift satellite has been
the presence, in approximately 30 per cent of the X-ray afterglows, of
flaring activity on top of the smooth afterglow decay (Burrows et
al. 2005; Falcone et al. 2006, 2007; Nousek 2006; Chincarini et al,
2007, 2010, Margutti et al. 2010b). The temporal properties of the
flares, and especially the fact that they have a duration that is
shorter than the time at which they appear in the light curve, have
lead to the conclusion that the flares' most likely origin is
late-time activity of the inner engine (Ioka et al. 2005; Zhang et
al. 2006; Lazzati \& Perna 2007, Curran et al. 2008). As a matter of
fact, any disturbance in the brightness that takes place
simultaneously on the surface of a spherical relativistic fireball
produces a flare whose duration is at least equal to the time of the
peak (see, however, Giannios 2006 for a model of magnetic dissipation
on the external shock). In addition, X-ray flares show spectral and
temporal properties similar to the prompt emission of GRBs (Margutti
et al. 2010a).

The inner engine of GRBs therefore has to be active, at least in some
cases, for a time much larger than the few hundreds of seconds of the
duration of the prompt emission (Kouveliotou et al. 1993). The
mechanism that keeps the engine alive and the mechanism for the
production of the flares are, however, still matter of active
debate. In the collapsar scenario (Woosley 1993), the reason for the
continued activity of the engine is either fall-back material from the
exploding star (Chevalier 1986; MacFadyen et al. 2001; Zhang \&
Woosley 2008) or the disk itself that takes some time to accrete all
its matter (Cannizzo et al. 1990). In the magnetar model (e.g. Usov
1992; Bucciantini et al. 2008), on the other hand, the engine activity
is due to the continuous spin-down of the neutron star. What is
unclear, for any choice of the inner engine, is the source of the
intermittent behavior of the flaring activity. While during the prompt
phase the engine is active for most of the time, during the flaring
phase the engine is mostly inactive.

\begin{figure*}
\centerline{\psfig{file=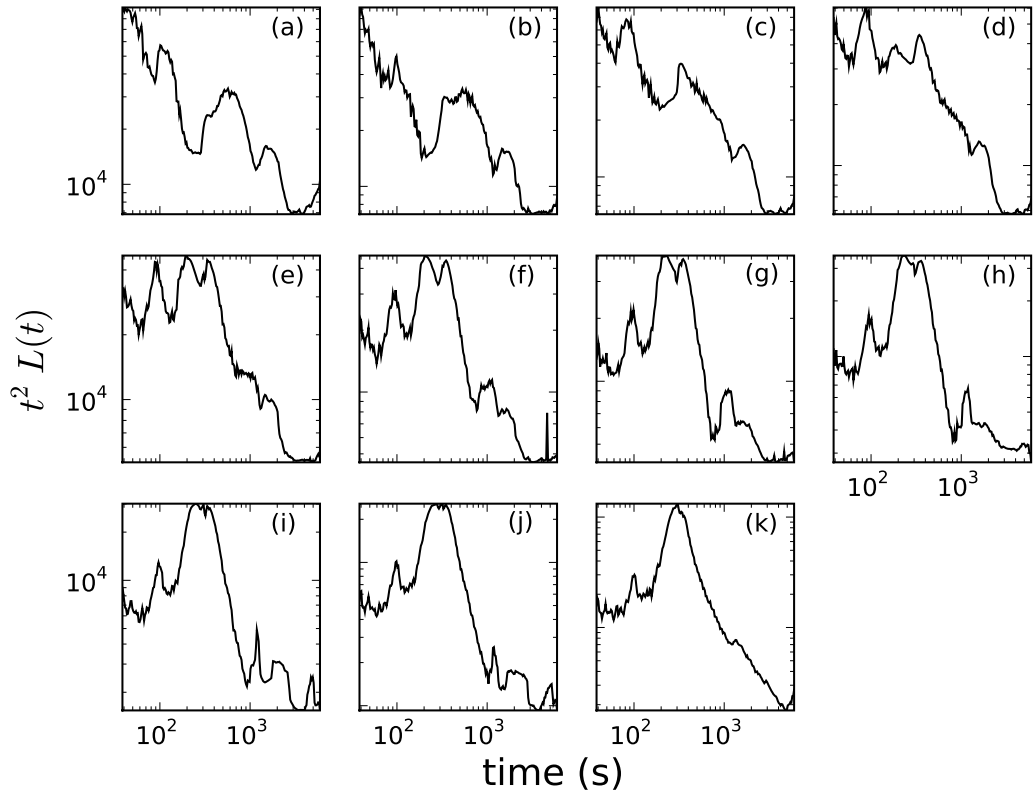,width=0.88\textwidth} }
\caption{Light-power curves derived from our numerical
  simulation. Light-power curves were calculated for observers lying
  at viewing angles of $1, 2, 3, 4, 5, 6, 7, 8, 9, 10$, and 16 degrees
  from the jet axis [panels (a) through (k)]. All the light-power
  curves are multiplied by $t^2$ to enhance the presence of flares on
  top of the overall decay behavior. }
\label{fig:lc}
\end{figure*}

Until now, models had concentrated on finding a source of variability
that is intrinsic to the engine, i.e., the engine itself releases
energy in a flaring fashion. Perna et al. (2006) proposed
gravitational instability in the outer parts of the accretion disk as
the source of the flare intermittent behavior (see also Proga \& Zhang
2006; Piro \& Pfahl 2007).  Lazzati et al. (2008) proposed that the
switch from a neutrino cooled thin disk to a thick disk could be the
reason for the change in behavior. In this paper, we show that even if
the inner engine is continuous with a featureless power-law decay, the
jet interaction with the disrupting star brings about variations in
the jet energy that resemble in luminosity and temporal evolution the
properties of the observed flares.

This paper is organized as follows: in Sect. 2 we describe our
numerical simulation and how we derived light curves, in Sect. 3 we
discuss the characterization of the flares and the results,  in
Sect. 4 we discuss a physical mechanism for the fluctuations, and in
Sect. 5 we discuss and summarize our findings.

\section{Numerical Simulation and light-power curves}

This paper is based on the results of a numerical simulation of a GRB
progenitor with a central engine that stays active for a long
time. The progenitor star used in the simulation was model 16TI from
Woosley \& Heger (2006), a 16 solar masses Wolf-Rayet star evolved to
pre-explosion. This progenitor is the same used in previous
simulations from our group (Morsony et al. 2007, 2010; Lazzati et
al. 2009, 2010). The jet was injected as a boundary condition with an
opening angle $\theta_0=10^\circ$, a Lorentz factor $\Gamma_0=5$ and
internal energy sufficient to reach an asymptotic Lorentz factor
$\Gamma_\infty=400$, in case of full, non-dissipative
acceleration. The jet luminosity was initially set to constant
$L_0=5.33\times10^{50}$~erg~s$^{-1}$. At 7.5~s after the ignition the
luminosity was changed into a smoothly decaying power-law
$L(t)=L_0(t/7.5)^{-5/3}$, where the exponent was chosen to mimic the
accretion rate of fall-back material\footnote{Note that the assumption
  of a linear relation between accretion rate and outflow luminosity
  is appropriate for magnetic jet driving (e.g. Barkov \& Komissarov
  2010) while a steeper relation should be used in case of a neutrino
  driven outflow ($L_{\rm{jet}}\propto\dot{m}^{9/4}$, Zalamea \&
  Beloborodov 2010).} (Chevalier 1989; MacFadyen et al.  2001). The
engine was turned off at 6000~s after ignition, at which time the
simulation was also stopped.  The simulation was performed with the
adaptive mesh refinement code FLASH (Fryxell et al. 2000) with
resolution and refinement scheme analogous to Morsony et al. (2007).

Our simulation has two major limitations that were made necessary by
technical limitations and by the need of simplifying the setup in
order to be able to extend the engine life-time maintaining the
overall run time reasonable. First, even though we mention fallback
accretion as a possible source of the extended energy release from the
inner engine, we do not simulate the actual fall-back inflow. Second,
our simulation does not include magnetic fields, given the technical
difficulty of performing special-relativistic magneto-hydrodynamic
simulations with an adaptive mesh refinement scheme. Despite these two
limitations we believe that our simulations give a credible
description of the jet-star interaction. The fall-back inflow should
not affect the jet dynamics in our computation box both because we do
not simulate the inner core of the star and because the inflow is
expected to be more relevant in the equatorial area. The effect of the
lack of magnetic fields is more difficult to evaluate. As a matter of
fact, it is very likely that the jet is launched by a magnetic
mechanism, especially during the low-luminosity phase when the
neutrino cooling has been switched-off. The subsequent evolution of
the jet during the acceleration phase decreases the magnetization and,
at the radii of the outer boundary of our computational box, it is
likely that the jet is already matter dominated (Tchekhovskoy et
al. 2010). Hopefully special relativistic MHD simulations with a
similar setup will become feasible in the near future to confirm the
details of our results.

The simulation was analyzed to derive light-power curves as
described in Morsony et al. (2010). Light-power curves are proxies to
the light curves that can be computed from simulations that do not
extend in radius far enough to allow for the direct comparison of the
light curve. Light-power curves assume that the efficiency of
conversion of energy into radiation is constant. Light-power curves
are not sensitive to spectral evolution, and are therefore a proxy to
the bolometric light curve and not to the light curve in a particular
band. The light-power curves for observers at angles of $1, 2, 3, 4,
5, 6, 7, 8, 9, 10$, and $16$ degrees were calculated, including only
material with an asymptotic Lorentz factor of at least 10. Light-power
curves were extracted at a radius of $2.5\times10^{11}$~cm, close to
the outer boundary of our simulation.

\section{Data Analysis and Results}

\begin{center}
\begin{figure}
\centerline{\psfig{file=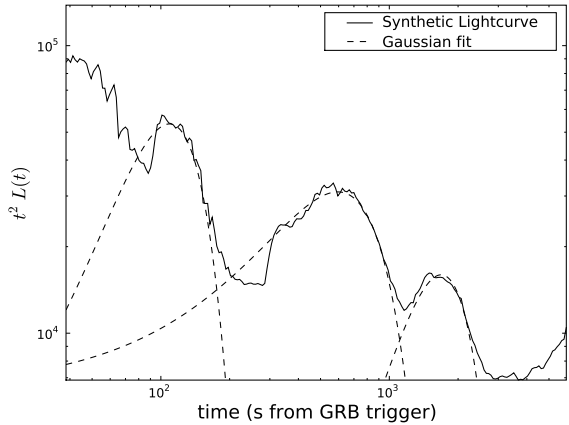,width=0.95\columnwidth}}
\caption{A sample light curve (for an observer at $\theta=1^\circ$),
  with best fit Gaussian functions overlaid on the identified flares.}
\label{fig:gauss}
\end{figure}
\end{center}
 
The light-power curves obtained from the simulation are shown in
Figure~\ref{fig:lc}, with the luminosity multiplied by a factor $t^2$
to emphasize the presence of bumps, the candidate flares.  A first
finding of our simulation is that the light-power curves decay in time
faster than the injected $t^{-5/3}$.  This is a consequence of the
increase in the opening angle of the jet as the confining power of the
progenitor star wanes (Lazzati \& Begelman 2005; Morsony et
al. 2007). We find that the slope in the time interval $200<t<1000$ s,
when the opening angle is growing, is roughly $L(t)\propto T^{-2.4}$,
flattening to roughly $t^{-2}$ at later time, when the opening angle
enters an almost constant phase. This is an important finding because
it highlights the fact that the steep slopes found in the average
flare luminosity ($t^{-1.5}$, Lazzati et al. 2008; or even $t^{-2.7}$,
Margutti et al. 2010b) may not reflect directly the behavior of the
inner engine, and a flatter input is required to reproduce the
observations. It is worth emphasizing here that the opening angle
$\theta_{\rm{sat}}\sim10^\circ$ at which the increase stops, and
consequently the time at which the decay flattens, is dictated by the
input opening angle of the simulation
($\theta_0\sim\theta_{\rm{sat}}$) and therefore we do not expect that
either the limiting opening angle or the time at which the flattening
is observed should be universal among GRBs.

In order to find and characterize the flares that are present in the
light-power curves we first subtracted a broken power-law
continuum. Typically the subtracted background had a steep early phase
and a flatter final, as discussed above for the overall light-power
curves. Flares were identified as positive fluctuations in the flux of
the background-subtracted curve and characterized by fitting a
Gaussian profile letting the peak time, normalization, and width of
the Gaussian free to vary. The fits were performed by minimizing the
$\chi^2$ statistics, assuming a uniform uncertainty in the data (see
Figure~\ref{fig:gauss} for a flare fitting example). Gaussian
functions were used in order to make our results directly comparable
to the observational results of Chincarini et al. (2007).  For the
identified flares the ratio of the flare peak flux to the continuum
flux under the flare was also evaluated.  It is worth noticing that
this ratio is an upper limit to what actual observations would provide
because we do not have any external shock component in our calculation
(see also Section 4).

\begin{figure}
\centerline{\psfig{file=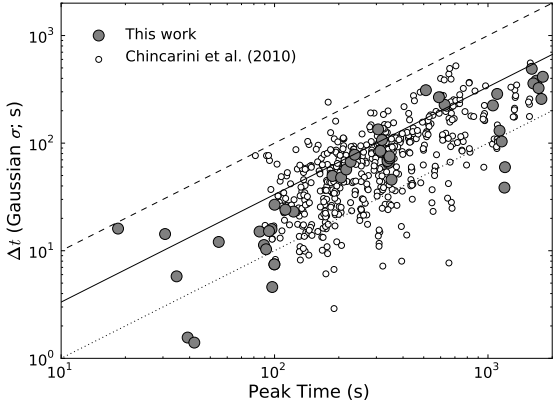,width=0.9\columnwidth}}
\caption{Results of the flare characterization from the synthetic
  light-power curves in the ($t$, $\Delta t$) plane. Each gray dot
  represents the Gaussian width and peak time of one identified
  flare. Open, smaller circles are the results of Swift X-ray flares
  characterization by Chincarini et al. (2010). The dashed, solid,
  and dotted lines represent the three levels $\Delta t/t=1, 0.3$, and
  $0.1$, respectively.}
\label{fig:dtt}
\end{figure}

\begin{figure}
\centerline{\psfig{file=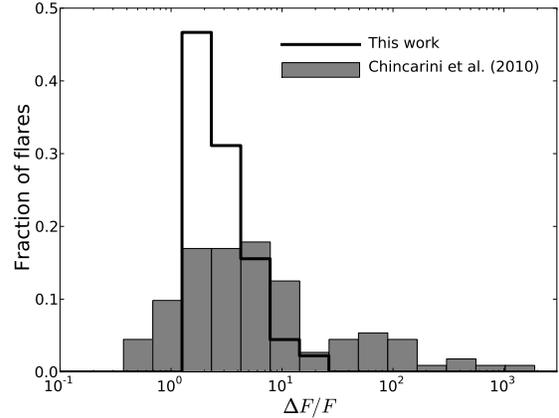,width=0.9\columnwidth}}
\caption{Histogram of the flare contrast from our simulation (thick
  solid line) compared to observations (shaded
  area).}
\label{fig:dff}
\end{figure}

The results of the flares characterization are shown in
Figure~\ref{fig:dtt} and Figure~\ref{fig:dff}. The simulation flares
are characterized by a duration that is smaller than their peak time,
as found in observational surveys (Chincarini et al. 2007, 2010). In
addition, the flare duration is correlated to the flare peak time,
another observational finding (see Fig. 10 of Chincarini et
al. 2007). The typical flare contrast is of the order of a few,
comparable to what found by Chincarini et al. (2007, 2010) and
Margutti et al. (2010b). However, our simulation was unable to
reproduce the rare but very bright flares observed in some Swift GRBs,
with peak flux contrasts up to a hundred or even a thousand.

\section{Analytical Model}

The simulation that we have presented shows that flaring activity with
the temporal characteristics observed in Swift light curves can be
reproduced in a simulation with a long-lasting engine that does not
have any intrinsic flaring behavior. The origin of the flares lies
therefore in propagation instabilities within the jet.

Any variation of the jet luminosity should be accompanied by a
variation in the jet opening angle. During the hot phase of the jet,
when the jet pressure is high, an increase of the jet
luminosity\footnote{Note that the jet luminosity can temporally
  increase even if the engine luminosity decreases. For example, a
  slower outflow ahead of a faster portion of the jet produces a
  pile-up of jet material that locally increases the jet luminosity
  without the need of an increase of the engine luminosity.}  should
be accompanied by an increase of the opening angle, since the
confining pressure of the star is uniform and the increased jet
pressure would cause an increase of the opening angle. On the other
hand, during the cold phase of the jet, a reduction of the opening
angle would bring about an increase of the jet luminosity (at least
the specific luminosity per unit solid angle that is responsible for
the observed light curve). We therefore expect that early flares are
accompanied by an opening angle increase while late flares are
simultaneous to a shrinking of the opening angle. This, at least,
should be true for the inner parts of the jet. For lines of sight that
are close to the edge of the jet, another important factor is whether
or not the jet opening angle crosses the line of sight. If the jet
opening angle grows to include a particular line of sight, the
corresponding light-power curve will show a prominent bump, while a
depression would be seen if the jet opening angle decreases. Light
curves for observers close to the jet axis and observers close to the
jet edge could therefore be anti-correlated. An example of this
behavior is the depression observed at $T\sim200$~s in panels (a),
(b), and (c) of Figure~\ref{fig:lc} (viewing angles of 1, 2, and 3
degrees). At the same time, a bump is seen in panels (e), (f), and
(g), for viewing angles of 5, 6, and 7 degrees.

\begin{figure}
\centerline{\psfig{file=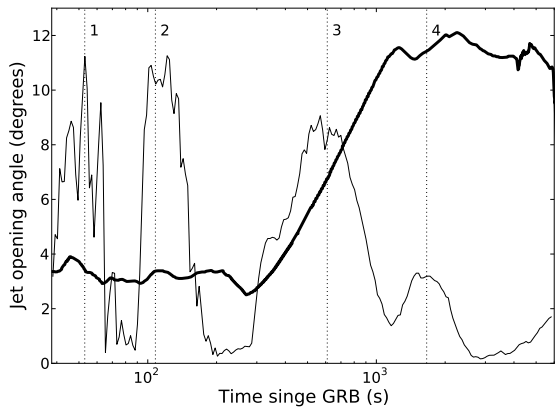,width=0.9\columnwidth}}
\caption{Comparison between the jet opening angle at
  $R=2.5\times10^{11}$~cm (thick solid line) and the background
  subtracted light-power curve at $\theta=1^\circ$ (thin solid
  line). Prominent flares in the light-power curve have been
  highlighted with vertical dotted lines and identified wit numbers 1,
  2, 3, and 4.}
\label{fig:lcan}
\end{figure}

To check for this occurrence, we have calculated the opening angle of
the jet to see if there are any variations of the jet opening angle in
coincidence with flares in the light curve. The result of this
analysis is shown in Figure~\ref{fig:lcan} for the light-power curve
calculated at $1^\circ$ off-axis. There are four prominent flares in
the curve and they have been identified with numbers 1, 2, 3, and
4. The comparison indeed reveals that early flares (1 and 2) are
accompanied by humps in the opening angle plot, while the late flare
(4) is accompanied by a depression. Flare 3 is difficult to evaluate,
since it lays on top of an overall growth of the opening angle of the
jet.

The duration of flares that are accompanied by a variation in the jet
opening angle can be computed assuming that the mechanisms that
activates and quenches them is the pressure of the surrounding stellar
material. In that case, the time it takes for the opening angle to
vary can be computed by knowing the jet transverse size and the sound
speed in the exploding stellar material.

The time to restore the jet opening angle after a factor of 2 increase
in kinetic luminosity (a $\sqrt{2}$ variation in opening angle) would
be
\begin{equation}
\delta t\simeq \frac{R_\perp}{3v_s}\simeq\frac{R_\star\theta_j}{3v_s}
\label{eq:dt}
\end{equation}
where $R_\perp$ is the jet transverse radius, $v_s$ is the sound speed
of the exploding star, $R_\star$ is the stellar radius, and $\theta_j$
is the jet opening angle. The sound speed of the star is:
\begin{equation}
  v_s\simeq\sqrt{\frac{kT_\star}{m_p}}
\label{eq:vs}
\end{equation}
where $T_\star$ is the average temperature of the star's material and
$m_p$ is the proton mass. The temperature of the exploding star can be
calculated assuming that the expansion is adiabatic and at constant
speed as:
\begin{equation}
  T_\star\simeq\frac{T_0 R_\star^2}{(v_{ej} \,t)^2}
\label{eq:tstar}
\end{equation}
where $T_0$ is the average stellar temperature immediately after the
explosion and $v_{ej}$ the average ejecta radial velocity. These two
quantities can be calculated as
\begin{equation}
  T_0\simeq\frac{2Em_p}{3kM_\star}
\label{eq:t0}
\end{equation}
and
\begin{equation}
  v_{ej}\simeq\sqrt{\frac{2E}{M_\star}}
\label{eq:vej}
\end{equation}
where $M_\star$ is the stellar mass and we have assumed that all the
explosive energy ($E$) goes into thermal motions (for Eq.~\ref{eq:t0})
or expansion velocity (for Eq.~\ref{eq:vej}). Combining all the above
equations, one can find a very simple solution for the typical flare
duration:
\begin{equation}
\frac{\delta t}{t}\simeq\frac{\sqrt{3}}{3} \theta\sim0.1
\label{eq:dttfinal}
\end{equation}
An important property of Eq.~\ref{eq:dttfinal} is that all the
dependence on the star property and explosive energy simplify, leaving
the opening angle of the jet as the only quantity that can alter the
typical flare duration. This simple model reproduces therefore two
fundamental properties of the observations: the fact that the flare
lasts less than the time at which it peaks and the fact that the flare
duration correlates linearly with the peak time.

\section{Discussion and Conclusions}

In summary, the flares that we observe in our simulation are due to
two different mechanisms. Early flares, when the jet propagation
through the star still excites turbulent motions (Morsony et al. 2007;
Mizuta et al. 2010), are due to velocity stratification within the
jet, and are accompanied by increases in the opening angle due to the
increased jet pressure. Late flares, instead, are due to reductions in
the jet opening angle. In both cases, as shown above, the flare
duration is smaller than but correlated to the flare peak time.

Even though our simulation shows flares with characteristics that are
similar to those observed, it does not show any strong flare with flux
contrast larger than about an order of magnitude. Extreme flares with
contrast of two or even three orders of magnitude have been observed
(Chincarini et al. 2010). Margutti et al. (2010b) show that the
distribution of flare contrasts is bimodal and two populations of
flares are present. The propagation instabilities that we present here
seem to be able to account for the small flares of the main peak (see
Figure~\ref{fig:dff} and Figure~7 in Margutti et al. 2010b). More
prominent flares likely require flaring activity in the central
engine. 

In addition to prominent flares in long-duration GRB afterglows, our
model cannot explain the presence of flares in the X-ray afterglows of
short-duration GRBs (e.g. Campana et al. 2006). Such flares need
therefore a source of variability in the inner engine, such as the
disk instability proposed by Perna et al. (2006). Our prediction is
therefore that the contrast distribution of short-duration GRB flares
should be analogous to the one of the brighter flares observed in
long-duration GRBs. Unfortunately, the flare sample from short GRB
afterglows is small and a quantitative comparison is not possible, at
this stage.

Another important finding of our simulation is that during the early
phases (roughly between few hundreds and one thousand seconds) the
light-power curve decay is much steeper than the input
luminosity. This is due to the fact that the opening angle of the jet
is increasing in that time span, causing the energy to be distributed
over a wider portion of the sky. Since the flares follow the
light-power curve behavior, the average flare luminosity (Lazzati et
al. 2008; Margutti et al. 2010b) also appears to be steeper than what
its source mechanism, without the spreading of the opening angle,
would produce. In this light, caution must be spent to interpret the
steep slopes measured in the average flare light curve since they may
be due to much shallower input luminosities. It is tantalizing that
Margutti et al. (2010) find a prominent flattening in the average
flare light curve at about 1000 s, right when the opening angle of the
jet stops growing in our simulation (even though, as mentioned before,
the time of the flattening in the simulation is dictated by the input
opening angle $\theta_0$ and by the evolution of the jet
luminosity. With a constant luminosity of the jet Morsony et al. 2010
find that the opening angle saturates much earlier at
$t\sim50$~s). Unfortunately, however, Margutti et al. (2010) also
found that flares at times larger than 1000 s are harder to robustly
characterize and their results should be taken with caution at long
times.

Finally, we would like to point out that our light-power curves do not
include any external shock component. Our flares are therefore likely
to be diluted by the synchrotron emission of the external shock. A
prediction of our model is therefore that the afterglows dominated by
the central engine (Ghisellini et al. 2009) should display stronger
flaring activity with respect to afterglow dominated by the external
shock component. An indication of this behavior is the decreased
occurrence of flares in simple afterglow with no breaks (see Fig. 16 of
Margutti et al. 2010a). A word of caution should also be spent for the
fact that our light-power curves are calculated at a distance of
$2.5\times10^{11}$~cm from the progenitor, close to the outer boundary
of our simulation. In reality, the radiation can only be released at
the jet photosphere or further out, at a distance at least one order
of magnitude larger than our extraction radius. It is possible that
the temporal structure of the flares is modified during the further
expansion of the jet making the flares longer in time and increasing
their $\Delta t/t$. Unfortunately, running a simulation for such a
long time (6000 s) forced us to use a relatively small box and made it
impossible to perform a fully self consistent calculation with an
extraction radius at the jet photosphere or beyond.

\section{Acknowledgments} 
We would like to thank Guido Chincarini and Raffaella Margutti for
kindly making their data available to us. This work was supported in
part by NASA ATP grant NNG06GI06G and Swift GI grant NNX08BA92G and by
the North Carolina State University Office of Undergraduate Research.

\end{document}